\documentclass[12pt]{iopart}
\usepackage{iopams}
\usepackage{graphicx}
\usepackage{hyperref}
\usepackage{booktabs}
\usepackage{xcolor}

\begin{document}

\title[Ginzburg-Landau Theory for Confined Thin-Film Superconductors]
{Ginzburg-Landau Theory for Confined Thin-Film Superconductors}

\author{G A Ummarino$^1$ and A Zaccone$^{2,3}$}

\address{$^1$ Dipartimento di Scienza Applicata e Tecnologia, Politecnico di Torino,
Corso Duca degli Abruzzi 24, 10129 Torino, Italy}

\address{$^2$ Department of Physics ``A. Pontremoli'', University of Milan,
via Celoria 16, 20133 Milan, Italy}

\address{$^3$ Institut f\"ur Theoretische Physik, University of G\"ottingen,
Friedrich-Hund-Platz 1, 37077 G\"ottingen, Germany}

\ead{giovanni.ummarino@polito.it}
\ead{alessio.zaccone@unimi.it}

\begin{abstract}
We develop a Ginzburg-Landau theory for superconducting thin films under quantum confinement. Starting from the Ginzburg-Landau theory and the recently developed confinement theory of metallic thin films, explicit analytical expressions are derived for the coherence length, penetration depth, electronic mean free path, and Ginzburg-Landau parameter in confined geometries.
The central result is that quantum confinement directly renormalizes the intrinsic superconducting coherence length through confinement-induced modifications of the electronic density of states and Fermi energy. This effect is absent in conventional thin-film transport theories based solely on surface scattering. As a consequence, confinement simultaneously suppresses the coherence length and enhances the penetration depth, thereby driving superconductors toward stronger type-II behavior over a broad thickness range. The interplay between quantum confinement and disorder can also produce reentrant type-I/type-II crossovers. The theory predicts a crossover regime in which confinement-induced renormalization of superconducting length scales and transport scattering become strongly intertwined. Comparison with recent penetration-depth measurements in Al thin films shows that the observed enhancement of the penetration depth originates from the interplay between confinement-induced renormalization of the coherence length and suppression of the effective mean free path by surface and disorder scattering. The results establish a direct connection between quantum confinement and superconducting electrodynamics in confined metallic films.
\end{abstract}

\vspace{2pc}
\noindent{\it Keywords}: Ginzburg-Landau theory, superconducting thin films, quantum confinement, penetration depth, coherence length, Fuchs-Sondheimer scattering

\maketitle

\section{Introduction}

The superconducting properties of metallic thin films undergo substantial modifications when the film thickness becomes comparable to characteristic electronic length scales such as the Fermi wavelength, the electron mean free path, and the superconducting coherence length. In this regime, quantum confinement alters the distribution of accessible electronic states in momentum space, modifies the density of states at the Fermi level, and changes the superconducting electrodynamics of the system. Thin superconducting films therefore represent an ideal platform where quantum size effects, electronic transport, and collective superconducting phenomena become strongly intertwined.

Understanding superconductivity in confined geometries is also of considerable technological importance. Thin-film superconductors are central components in superconducting quantum circuits, microwave resonators, kinetic-inductance detectors, and nanoscale superconducting electronics \cite{Blais2021,Brooks2023,Forn2025}. In these systems, superconducting length scales such as the penetration depth $\lambda(T)$ and the coherence length $\xi(T)$ directly control kinetic inductance, vortex behavior, magnetic screening, and device coherence properties. In particular, thin-film aluminum has become one of the most important materials in superconducting quantum technologies because of its reproducibility, compatibility with Josephson-junction fabrication, and low microwave losses \cite{FornDiaz2010,Blais2021}.

Experimentally, it has long been known that reducing the film thickness strongly enhances the superconducting penetration depth and may drive materials such as Al from type-I to type-II superconductivity \cite{Tinkham,deGennes}. Recent high-precision measurements by López-Núñez \textit{et al.} \cite{Forn2025} demonstrated a pronounced increase of the penetration depth in Al thin films as the thickness decreases from approximately $200\,\mathrm{nm}$ to below $30\,\mathrm{nm}$, together with signatures of a crossover between type-I and type-II superconducting behavior. Their work combined resonator measurements and transport measurements to extract the penetration depth with high accuracy and showed that conventional thin-film electrodynamics based solely on surface scattering and dirty-limit corrections is insufficient to provide a complete microscopic interpretation of the observed behavior.

In previous theoretical studies \cite{ref6,ref7}, the three-dimensional time-dependent Ginzburg--Landau equations were solved numerically for mesoscopic thin-film superconductors. These calculations established the relationship between the film thickness $L$ and the Ginzburg--Landau parameter $\kappa$, and identified the critical thickness at which the transition between type-I and type-II superconductivity occurs. Finally, for multiband superconductors, the crossover between type-I and type-II superconductivity has also been investigated using a perturbative approach \cite{ref8}.

In conventional approaches, the thickness dependence of superconducting properties is usually described through phenomenological transport models such as the Fuchs--Sondheimer theory and the Mayadas--Shatzkes framework \cite{Fuchs1938,Sondheimer1952,Mayadas1970}. Recent first-principles work has also emphasized that electron--surface scattering in nanoscale conductors can be treated microscopically and is strongly controlled by the electronic structure and surface properties \cite{ZhangLiu2024}. Within these descriptions, thickness primarily affects the electronic mean free path through diffuse surface scattering and grain-boundary scattering. The superconducting penetration depth then changes indirectly through dirty-limit corrections \cite{Tinkham,deGennes}. While this approach successfully captures part of the experimental phenomenology, it leaves open a fundamental question: whether quantum confinement itself can directly renormalize intrinsic superconducting length scales independently of disorder scattering.

Several recent works \cite{Travaglino2023,Ummarino2025,Zaccone2025,UmmarinoZaccone2024PRM,UmmarinoZaccone2025Mg,UmmarinoZaccone2026JPCM} introduced an analytical confinement theory for metallic thin films based on the geometric reconstruction of the Fermi surface under strong confinement. The theory avoids treating $k_z$ as a good quantum number in view of the ubiquitous atomic-scale disorder of the thin-film surface \cite{Zaccone2025,Mirigliano2021}, and predicts confinement-induced forbidden regions in momentum space together with a crossover between weak- and strong-confinement regimes. The theory has been shown to be able to quantitatively predict the experimentally observed trend of the superconducting critical temperature $T_c$ as a function of film thickness $L$ \cite{Ummarino2025}, typically with a maximum in $T_c$ at thickness value $L_c$ corresponding to the onset of the topological reconstruction of the Fermi surface (and coinciding with the crossover from weak to strong confinement). 

One of the most important consequences of this framework is a confinement-induced renormalization of the density of states and consequently of the electron-phonon coupling constant and Fermi energy, which in turn modifies superconducting observables such as the critical temperature.

More generally, earlier developments in multiband and confined superconductivity have highlighted the importance of multiple superconducting length scales and unconventional electrodynamics in reduced-dimensional systems~\cite{Babaev2011,SilaevBabaev2011}. In particular, Babaev and collaborators showed that the interplay between different coherence lengths and penetration depths can generate qualitatively new superconducting regimes beyond the standard type-I/type-II classification. Related work by Bianconi and collaborators emphasized the importance of shape resonances, Lifshitz transitions, and quantum confinement effects in superconducting nanofilms and multiband superconductors~\cite{Bianconi2005,Cariglia2016}. These developments further motivate the formulation of a confinement-renormalized Ginzburg--Landau theory in which superconducting length scales become explicit functions of confinement geometry and electronic structure. 

However, a systematic Ginzburg--Landau description of superconducting thin films derived from this confinement framework is still lacking. In particular, no theory presently exists, which explicitly connects confinement-induced modifications of the density of states and Fermi surface topology to the experimentally observed evolution of the penetration depth, coherence length, and Ginzburg--Landau parameter in thin superconducting films. The recent penetration-depth measurements of thin-film Al by López-Núñez \textit{et al.} therefore provide a particularly important experimental motivation for developing such a framework.

Further motivation comes from recent work on confined and disordered aluminum systems. Deshpande \textit{et al.} showed that granular Al thin films exhibit a superconducting dome whose height and position depend on deposition conditions and film thickness, with $T_c$ reaching values well above bulk Al~\cite{Deshpande2025}. Related theoretical work has also shown that superconductivity in metallic thin films can be strongly modified by external electric fields, with quantum confinement, electrostatic screening, and Eliashberg corrections playing an essential role~\cite{ZacconeFomin2024,ZacconeUmmarino2025PRB,UmmarinoZaccone2025PRM}. These developments support the need for a Ginzburg--Landau formulation in which confinement modifies not only transport parameters, but also intrinsic superconducting length scales.

The purpose of the present work is to formulate a complete Ginzburg--Landau theory for superconducting thin films starting from the microscopic confinement theory developed in Refs.~\cite{Travaglino2023,Zaccone2025}. The resulting framework provides explicit analytical expressions for the confinement-renormalized Ginzburg--Landau coefficients, coherence length, penetration depth, mean free path, and Ginzburg--Landau parameter. In contrast to conventional thin-film transport theories, the present approach predicts that quantum confinement directly renormalizes the intrinsic superconducting coherence length through the confinement dependence of the electronic density of states and Fermi energy.

The theory further provides a natural framework to interpret recent experimental observations in Al thin films, where the measured enhancement of the penetration depth appears to originate from the interplay between transport scattering and confinement-induced modifications of the intrinsic superconducting length scales. The present work therefore extends recent confinement theories of nanoscale transport and superconductivity into the Ginzburg--Landau regime and provides a unified phenomenological description connecting quantum confinement, electronic structure, disorder scattering, and superconducting electrodynamics in confined metallic films.

\section{Ginzburg--Landau Free Energy}

The Ginzburg--Landau superconductive free-energy density, for homogeneous systems, close to the superconducting transition temperature, is written as \cite{Tinkham,deGennes}

\begin{equation}
F_s(T)=F_n+\alpha(T)|\psi(T)|^2+\frac{b}{2}|\psi(T)|^4,
\end{equation}

where $F_n$ is the normal-state free energy, $\psi$ is the superconducting order parameter and the coefficients $\alpha(T)$ and $b$ are determined microscopically.

$\alpha(T)$ has to change sign at the critical temperature so one writes

\begin{equation}
\alpha(T)=\alpha_0\left(\frac{T-T_c}{T_c}\right).
\end{equation}

Within microscopic BCS theory the coefficient $\alpha_0$ is proportional to the density of states at the Fermi level \cite{Tinkham,deGennes},

\begin{equation}
\alpha_0=N(0),
\end{equation}

while the quartic coefficient is given by

\begin{equation}
b=\frac{7\zeta(3)}{8\pi^2}\frac{N(0)}{(k_B T_c)^2}.
\end{equation}

The superconductive free energy has a minimum when $\psi(T)=\psi_0(T)$
where
\begin{equation}
\psi_0(T)=\sqrt{|\frac{\alpha(T)}{\beta}|},
\end{equation}
and $\psi^2_0(T)$ represents the density of Cooper pairs.

The density of states $N(0)$ therefore constitutes the central quantity through which confinement modifies the superconducting properties.

\section{Quantum Confinement and Density of States}

In the confinement theory of thin metallic films, confinement along one spatial direction suppresses low-energy states in momentum space. The resulting modification of the available Fermi volume changes the density of states and shifts the Fermi energy.

The density of states becomes confinement dependent,

\begin{equation}
N_{\mathrm{film}}(0)=C(L)N_{\mathrm{bulk}}(0),
\end{equation}

where $C(L)$ is a confinement renormalization factor.

For weak confinement, corresponding to $L>L_c$, the confinement factor is

\begin{equation}
C(L)=\left(1+\frac{2}{3}\frac{\pi}{nL^3}\right)^{1/3},
\end{equation}

where $n$ is the electronic carriers density and $L_c=(\frac{2\pi}{n})^{1/3}$ is the crossover thickness.

In this regime one typically has

\begin{equation}
1\leq C\lesssim 1.1.
\end{equation}

Hence the density of states is only moderately enhanced.

For strong confinement, namely $L<L_c$, the Fermi surface undergoes a topological reconstruction (cf. Fig. 2 in Ref. \cite{Travaglino2023}) and the confinement factor becomes

\begin{equation}
C(L)=\frac{2}{6^{1/3}}\sqrt{\frac{L}{L_c}}.
\end{equation}

Therefore

\begin{equation}
0\leq C\leq 1.
\end{equation}

The corresponding Fermi energy is renormalized according to

\begin{equation}
\varepsilon_{F,\mathrm{film}}=C^2\varepsilon_{F,\mathrm{bulk}}.
\end{equation}

This relation follows directly from the confinement-induced reconstruction of the occupied momentum-space volume.

\section{Electronic Mean Free Path}

The electronic mean free path is written as

\begin{equation}
\ell=v_F\tau,
\end{equation}

where $\tau$ is the scattering time and the Fermi velocity is

\begin{equation}
 v_F=\sqrt{\frac{2\varepsilon_F}{m}}.
\end{equation}

As a first approximation, we assume that the electronic scattering time
$\tau$ remains approximately independent of film thickness. This
assumption isolates the intrinsic effect of quantum confinement on the
electronic mean free path through the confinement-induced renormalization
of the Fermi velocity, independently of additional thickness-dependent
transport mechanisms. Under this approximation one finds
\begin{equation}
\ell_{\mathrm{film}}
=
\ell_{\mathrm{bulk}}
\sqrt{\frac{\varepsilon_{F,\mathrm{film}}}{\varepsilon_{F,\mathrm{bulk}}}}.\label{ellfilm}
\end{equation}
Equation~\ref{ellfilm} therefore represents the intrinsic (or ``bare'')
confinement contribution to the mean free path. In realistic thin films,
however, the scattering time also becomes thickness dependent because of
diffuse surface scattering, grain boundaries, interface roughness, and
other disorder effects. These additional transport mechanisms are not
included in the present first-order derivation and are incorporated
later, in Sec.~\ref{sec:comparison}, through the effective mean free path
of Eq.~\ref{eq:leff}. Equation~\ref{eq:leff} should therefore be
regarded as an extension of the intrinsic confinement scaling derived
here rather than as a replacement of Eq.~\ref{ellfilm}.

Using

\begin{equation}
\varepsilon_{F,\mathrm{film}}=C^2\varepsilon_{F,\mathrm{bulk}},
\end{equation}

we obtain

\begin{equation}
\ell_{\mathrm{film}}=C\,\ell_{\mathrm{bulk}}.
\end{equation}

The electronic mean free path therefore scales linearly with the confinement factor.

\section{Coherence Length}

Within Ginzburg--Landau theory, close to critical temperature, the superconducting coherence length is defined as \cite{Tinkham,deGennes}

\begin{equation}
\xi(T)=\sqrt{\frac{\hbar^2}{2m|\alpha(T)|}}.
\end{equation}

Substituting the temperature dependence of $\alpha(T)$ gives

\begin{equation}
\xi(T)=\frac{\hbar}{\sqrt{2m|\alpha_0|}}\frac{T_c}{T-T_c}|^{1/2}=\xi(0)|\frac{T_c}{T-T_c}|^{1/2}.
\end{equation}

where

\begin{equation}
\xi(0)=\frac{\hbar}{2\sqrt{m\alpha_0}}.
\end{equation}

Since

\begin{equation}
\alpha_0\propto N(0),
\end{equation}

one immediately obtains the confinement-renormalized coherence length:

\begin{equation}
\xi_{\mathrm{film}}
=
\xi_{\mathrm{bulk}}
\sqrt{\frac{N_{\mathrm{bulk}}(0)}{N_{\mathrm{film}}(0)}}.
\end{equation}

Using the confinement factor,

\begin{equation}
\xi_{\mathrm{film}}=\frac{\xi_{\mathrm{bulk}}}{\sqrt{C}}.
\end{equation}

Hence confinement tends to reduce the coherence length whenever the density of states is enhanced.

In the weak-confinement regime the effect remains moderate because $C$ differs only weakly from unity. Near the crossover thickness $L_c$, however, the suppression becomes stronger.

In real polycrystalline thin films, reducing the film thickness usually
leads to a reduction of the grain size and therefore to an additional
suppression of the electronic mean free path. In the dirty limit, the
effective superconducting coherence length is determined jointly by the
intrinsic coherence length and by impurity scattering, following the
well-known interpolation discussed by Tinkham and employed in the recent
analysis of thin-film Al penetration-depth measurements
\cite{Tinkham,Forn2025}. This interpolation is not derived within the present confinement framework but adopted as the standard phenomenological dirty-limit relation widely used in the superconductivity literature \cite{Tinkham,Forn2025}. In bulk materials, the effective coherence
length is therefore given by
\begin{equation}
\frac{1}{\xi_{\mathrm{bulk,eff}}}
=
\frac{1}{\xi_{\mathrm{bulk}}}
+
\frac{1}{\ell_{\mathrm{bulk}}}.\label{matth}
\end{equation}

Applying the confinement-induced renormalization to Eq. \ref{matth} of both the
intrinsic coherence length and the electronic mean free path immediately
yields the corresponding expression for the confined film:
\begin{equation}
\frac{1}{\xi_{\mathrm{film,eff}}}
=
\frac{\sqrt{C}}{\xi_{\mathrm{bulk}}}
+
\frac{1}{C\,\ell_{\mathrm{bulk}}}.
\end{equation}

\section{London Penetration Depth}
Within Ginzburg--Landau theory, close to critical temperature, the superconducting London penetration depth, in the clean limit, is defined as \cite{Tinkham,deGennes}
\begin{equation}
\lambda_L(T)=\sqrt{\frac{2m}{4\mu_0e^2\psi^2_0(T)}},
\end{equation}
where $m$ is the electronic mass and $e$ the electronic charge. This definition can be written in another way:
\begin{equation}
\lambda_L(T)=\sqrt{\frac{m\beta}{2\mu_0e^2|\alpha(T)|}},
\end{equation}
but $\beta \propto T_c^{-2}$ so $\lambda_L \propto T_c^{-1}$ while the dependence from $N(0)$ disappears because is both in $\beta$ and in $\alpha$.

In the dirty limit the definition of the penetration depth is \cite{Forn2025}
\begin{equation}
\lambda_{\mathrm{bulk}}=\lambda_{L,\mathrm{bulk}}\sqrt{1+\frac{\xi_{\mathrm{bulk}} (T)}{\ell_{\mathrm{bulk}}(T)}},
\end{equation}

Applying the confinement-renormalized coherence length and mean free path derived above gives
\begin{equation}
\lambda_{\mathrm{film}}
=
\lambda_L
\left(\frac{T_{c,\mathrm{film}}}{T_{c,\mathrm{bulk}}}\right)
\sqrt{1+\frac{\xi_{\mathrm{bulk}}}{C\sqrt{C}\ell_{\mathrm{bulk}}}}.
\end{equation}

This expression contains two competing effects.

The first factor reflects the confinement dependence of the superconducting energy scale through the critical temperature. The second factor reflects the confinement-induced reduction of the electronic mean free path.

For weak confinement, namely $L>L_c$, the confinement factor remains close to unity. Consequently the penetration depth increases only moderately as the thickness approaches the crossover thickness from above.

In contrast, in the strong-confinement regime $C\rightarrow 0$ as $L\rightarrow 0$, which implies

\begin{equation}
\sqrt{1+\frac{\xi_{\mathrm{bulk}}}{C\sqrt{C}\ell_{\mathrm{bulk}}}}
\rightarrow \infty.
\end{equation}

At the same time, however,

\begin{equation}
\frac{T_{c,\mathrm{film}}}{T_{c,\mathrm{bulk}}}
\rightarrow 0
\end{equation}

may suppress the overall penetration depth depending on the asymptotic behavior of $T_c$.

This competition between the divergence induced by the mean free path correction and the suppression induced by the superconducting energy scale represents one of the key physical features of the strong-confinement regime.

\section{Ginzburg--Landau Parameter}

The Ginzburg--Landau parameter is defined as

\begin{equation}
\kappa=\frac{\lambda}{\xi}.
\end{equation}

Substituting the confinement-renormalized expressions for the penetration depth and coherence length yields
\begin{equation}
\kappa_{\mathrm{film}}
=
\kappa_{\mathrm{bulk}}
\left(\frac{T_{c,\mathrm{film}}}{T_{c,\mathrm{bulk}}}\right)\sqrt{C}
\left(1+\frac{\xi_{\mathrm{bulk}}}{C\sqrt{C}\ell_{\mathrm{bulk}}}\right)^{1.5}
\end{equation}

In the weak-confinement regime the Ginzburg--Landau parameter increases as the thickness approaches $L_c$ from above. Quantum confinement therefore intrinsically drives superconductors toward stronger type-II behavior over a broad thickness range, although the interplay between confinement and disorder may produce a reentrant type-I/type-II crossover in the strong-confinement regime. This prediction becomes especially important in nanometric films and superconducting nanodevices, where confinement-induced renormalization of superconducting length scales can qualitatively modify vortex physics, magnetic screening, kinetic inductance, and the electrodynamic response of the superconducting condensate.

\section{Explicit Thickness Dependence of Physical Quantities}

 The confinement factor is
\begin{equation}
C(L)\equiv \frac{N_{\rm film}(0)}{N_{\rm bulk}(0)}.
\end{equation}

For \(L>L_c\),
\begin{equation}
C(L)=\left(1+\frac{2\pi}{3nL^3}\right)^{1/3},
\end{equation}
whereas for \(L<L_c\),
\begin{equation}
C(L)=\frac{2}{6^{1/3}}\left(\frac{L}{L_c}\right)^{1/2}.
\end{equation}

The Fermi energy is therefore
\begin{equation}
\varepsilon_F(L)=
\left\{
\begin{array}{ll}
\varepsilon_F^{\rm bulk}
\left(1+\frac{2\pi}{3nL^3}\right)^{2/3},
& L>L_c,\\[1.0em]
\frac{\hbar^2}{m}
\left[\frac{(2\pi)^3 n}{L}\right]^{1/2},
& L<L_c.
\end{array}
\right.
\end{equation}

The density of states at the Fermi level is
\begin{equation}
N_{\rm film}(0)=C(L)N_{\rm bulk}(0).
\end{equation}

Thus, for \(L>L_c\),
\begin{equation}
N_{\rm film}(0)=N_{\rm bulk}(0)
\left(1+\frac{2\pi}{3nL^3}\right)^{1/3},
\end{equation}
while for \(L<L_c\),
\begin{equation}
N_{\rm film}(0)=N_{\rm bulk}(0)
\frac{2}{6^{1/3}}\left(\frac{L}{L_c}\right)^{1/2}.
\end{equation}

The superconducting critical temperature follows from the BCS expression
\begin{equation}
k_B T_c=1.13\,\varepsilon_D
\exp\left[-\frac{1}{U N_{\rm film}(0)}\right].
\end{equation}
where $\varepsilon_D$ is the Debye energy and $U$ is connected with the electron-phonon interaction.
Therefore,
\begin{equation}
T_c(L)=
\left\{
\begin{array}{ll}
T_D
\exp\left[
-\frac{1}{U N_{\rm bulk}(0)}
\left(1+\frac{2\pi}{3nL^3}\right)^{-1/3}
\right],
& L>L_c,\\[1.2em]
T_D
\exp\left[
-\frac{6^{1/3}}{2U N_{\rm bulk}(0)}
\left(\frac{L_c}{L}\right)^{1/2}
\right],
& L<L_c.
\end{array}
\right.
\end{equation}
where
\begin{equation}
T_D=\frac{1.13\,\varepsilon_D}{k_B}.
\end{equation}
is the Debye temperature.

The Ginzburg--Landau coefficient becomes
\begin{equation}
\alpha_0(L)=C(L)\alpha_{\mathrm{0,bulk}}.
\end{equation}

The coherence length is
\begin{equation}
\xi(L)=\frac{\xi_{\rm bulk}}{\sqrt{C(L)}}.
\end{equation}

Hence,
\begin{equation}
\xi(L)=
\left\{
\begin{array}{ll}
\xi_{\rm bulk}
\left(1+\frac{2\pi}{3nL^3}\right)^{-1/6},
& L>L_c,\\[1.2ex]
\xi_{\rm bulk}
\left[
\frac{6^{1/3}}{2}
\left(\frac{L_c}{L}\right)^{1/2}
\right]^{1/2},
& L<L_c.
\end{array}
\right.
\end{equation}

and in the diffusive limit
\begin{equation}
\frac{1}{\xi_{\mathrm{eff}}}=\frac{\sqrt{C}}{\xi_{\mathrm{bulk}}}+\frac{1}{C\,\ell_{\mathrm{bulk}}}.
\end{equation}

The mean free path scales as
\begin{equation}
\ell(L)=C(L)\ell_{\rm bulk}.
\end{equation}

Therefore,
\begin{equation}
\ell(L)=
\left\{
\begin{array}{ll}
\ell_{\rm bulk}
\left(1+\frac{2\pi}{3nL^3}\right)^{1/3},
& L>L_c,\\[1.2em]
\ell_{\rm bulk}
\frac{2}{6^{1/3}}
\left(\frac{L}{L_c}\right)^{1/2},
& L<L_c.
\end{array}
\right.
\end{equation}

The dirty-limit penetration depth is
\begin{equation}
\lambda(L)=\lambda_L(L)
\sqrt{1+\frac{\xi(L)}{\ell(L)}}.
\end{equation}

Using the simplified scaling
\begin{equation}
\lambda_L(L)\simeq \lambda_{L,\rm bulk}
\frac{T_c(L)}{T_{c,\rm bulk}},
\end{equation}
one obtains
\begin{equation}
\lambda(L)=
\lambda_{L,\rm bulk}
 \exp\left[
\frac{1}{U N_{\rm bulk}(0)}
\left(
\frac{C(L)-1}{C(L)}
\right)
\right]
\sqrt{
1+
\frac{\xi_{\rm bulk}}
{C(L)\sqrt{C(L)}\,\ell_{\rm bulk}}
}.
\end{equation}

Finally, the Ginzburg--Landau parameter becomes
\begin{eqnarray}
\kappa(L)
&=&
\frac{\lambda}{\xi_{\rm eff}}
\nonumber\\
&=&
\kappa_{\rm bulk}
\exp\left[
\frac{1}{U N_{\rm bulk}(0)}
\left(
\frac{C(L)-1}{C(L)}
\right)
\right]
\sqrt{C(L)}
\nonumber\\
&&\times
\left[
1+
\frac{\xi_{\rm bulk}}
{C(L)\sqrt{C(L)}\,\ell_{\rm bulk}}
\right]^{3/2}.\label{GLparam}
\end{eqnarray}

\section{Discussion}

The present framework provides a unified description of the superconducting properties of confined metallic thin films. The theory directly connects the microscopic confinement-induced reconstruction of the Fermi surface to macroscopic superconducting observables.

Several general trends emerge naturally from the theory:

(i) confinement enhances the density of states at the Fermi level in the weak-confinement regime, thereby increasing the superconducting critical temperature;

(ii) the coherence length decreases because the Ginzburg--Landau coefficient $\alpha_0$ scales with the density of states;

(iii) the penetration depth increases due to the reduction of the electronic mean free path.

Finally, the Ginzburg--Landau parameter increases as the film thickness approaches the crossover thickness, implying stronger type-II behavior in confined geometries. As discussed below, the interplay between confinement and disorder may produce a reentrant type-I/type-II crossover in the strong-confinement regime.

The strong-confinement regime is especially interesting because the Fermi surface ceases to remain spherical and undergoes a topological reconstruction, as demonstrated in Ref. \cite{Travaglino2023}, cf. Fig. 2 therein. In this regime, the standard isotropic Fermi-liquid picture may require anisotropic corrections and nonlocal electrodynamics.

Future developments should include Eliashberg strong-coupling corrections, disorder effects, anisotropic pairing and vortex physics in confined geometries.

\section{Comparison with Experimental Data}
\label{sec:comparison}
In order to assess the validity of the proposed framework, the theory was compared against experimental measurements of the penetration depth in superconducting Al thin films reported by Forn-Díaz and collaborators ~\cite{Forn2025}.

A first attempt based solely on confinement-induced renormalization of the density of states proved insufficient to reproduce the experimental trend. In particular, for Al films in the thickness range $20$--$200\,\mathrm{nm}$, the confinement correction factor $C(L)$ remains extremely close to unity because the confinement crossover thickness is subnanometric. Therefore, in this thickness range the dominant effect controlling the penetration depth is not the direct modification of the density of states, but rather the thickness dependence of the electronic mean free path through the residual resistivity.

Following the same philosophy employed in the Fuchs--Sondheimer (FS) formalism and in recent confinement-resistivity theories, the resistivity contributions were combined through Matthiessen's rule,
\begin{equation}
\rho(L)=\rho_{\mathrm{FS}}(L)+\rho_{\mathrm{dis}}(L),
\end{equation}
where $\rho_{\mathrm{FS}}$ is the Fuchs--Sondheimer surface-scattering contribution and $\rho_{\mathrm{dis}}$ is an additional disorder or grain-boundary contribution. This decomposition follows the same general strategy recently developed for confined transport in ultrathin films, where the resistivity is modeled as the sum of a FS surface-scattering contribution and a confinement-induced contribution associated with suppression of electronic states under confinement~\cite{ZacconePRM2025,ZacconeET2026}.

For the present Al thin films, we employ the reduced effective carrier density previously introduced in the Eliashberg analysis of Ummarino and Zaccone~\cite{Ummarino2025}, which yields a confinement crossover thickness \(L_c\simeq16.9\,\mathrm{nm}\). Consequently, the thinnest experimental samples lie close to the crossover regime where confinement-induced renormalization of the superconducting coherence length becomes observable. In this regime, confinement affects the superconducting electrodynamics both through the intrinsic renormalization of the superconducting length scales and through the suppression of the effective mean free path by surface and disorder scattering.

The FS correction reads
\begin{equation}
\frac{\rho_{\mathrm{FS}}(L)}{\rho_0}
=
1+
\frac{3}{8}(1-p)\frac{\ell_{\rm bulk}}{L},
\end{equation}
where $p$ is the surface specularity parameter.

The effective mean free path is therefore written as
\begin{equation}
\ell_{\mathrm{eff}}(L)
=
\frac{C(L)\ell_{\rm bulk}}{
1+
\frac{3}{8}(1-p)\frac{\ell_{\rm bulk}}{L}
+
\gamma\frac{\ell_{\rm bulk}}{L}
},\label{eq:leff}
\end{equation}
where the additional parameter $\gamma$ phenomenologically captures disorder and grain-boundary scattering effects which become increasingly important in ultrathin films.

The penetration depth is then modeled as
\begin{equation}
\lambda(L)
=
\lambda_{L,\rm bulk}
\cdot \exp\left[\frac{1}{UN_{\rm bulk}(0)}\left(\frac{C(L)-1}{C(L)}\right)\right]
\sqrt{
1+
\frac{\xi(L)}{\ell_{\mathrm{eff}}(L)}
},
\end{equation}
where the confinement-renormalized coherence length is given by
\begin{equation}
\xi_(L)=\frac{\xi_{\mathrm{bulk}}}{\sqrt{C(L)}}.
\end{equation}
We choose $\xi_{\mathrm{bulk}}=1600\,\mathrm{nm}$ and $\lambda_L=15.7\,\mathrm{nm}$ for Al. The value $\lambda_L=15.7\,\mathrm{nm}$ corresponds to the clean London penetration depth entering the dirty-limit expression, whereas the larger experimentally observed penetration depth already includes scattering and electrodynamic corrections. In the weak-coupling limit~\cite{Carbotte1990},
\begin{equation}
UN_{\mathrm{bulk}}(0)=\frac{\lambda-\mu^{*}}{1+\lambda},
\end{equation}
where, here, $\lambda=0.43$ is the electron--phonon coupling constant and $\mu^{*}=0.143$ is the Coulomb pseudopotential~\cite{Ummarino2025}, yielding
\begin{equation}
UN_{\mathrm{bulk}}(0)\approx0.20.
\end{equation}
The other parameters appearing in the formulas are free and will be determined by the fit.

The fit to the experimental data is shown in Fig.~\ref{fig:fit}.

\begin{figure}[ht]
\centering
\includegraphics[width=0.75\textwidth]{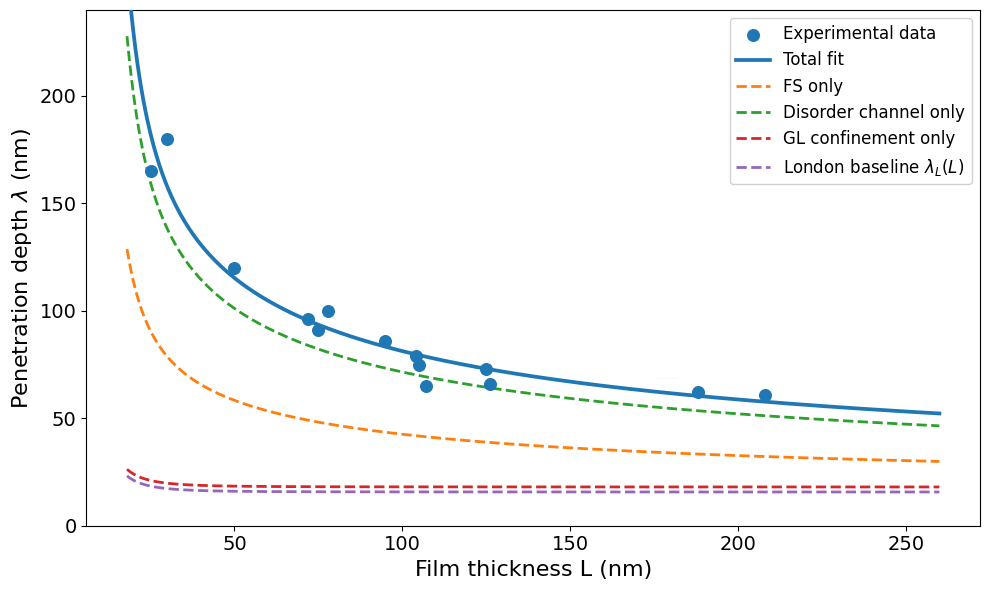}
\caption{
Comparison between the experimental penetration-depth data for Al thin films and the theoretical model. The symbols denote the experimental data of Ref.~\cite{Forn2025}, while the solid line is the full fit including Fuchs--Sondheimer surface scattering, an additional disorder/grain-boundary scattering channel, and the confinement-induced Ginzburg--Landau renormalization of the coherence length. The dashed curves show the partial contributions obtained by retaining only the Fuchs--Sondheimer channel, only the disorder channel, only the direct Ginzburg--Landau confinement correction, and the London baseline $\lambda_L(L)$. The effective carrier density is chosen consistently with the Eliashberg analysis of Al thin films in Ref.~\cite{Ummarino2025}, giving $L_c\simeq16.9\,\mathrm{nm}$. With this choice, the confinement contribution to $\xi(L)$ becomes visible near the thinnest films, although the dominant enhancement of $\lambda(L)$ still arises from the suppression of the effective mean free path by surface and disorder scattering.
}
\label{fig:fit}
\end{figure}

The fit parameters are summarized in Table~\ref{tab:fit_parameters}. In this comparison we use the reduced effective carrier density $n_{\mathrm{eff}}=1.30\times10^{24}\,\mathrm{m^{-3}}$, consistent with the Eliashberg analysis of Al thin films by Ummarino and Zaccone~\cite{Ummarino2025}. This choice gives a confinement crossover thickness $L_c\simeq16.9\,\mathrm{nm}$, bringing the thinnest samples closer to the regime where the confinement-induced renormalization of the superconducting coherence length becomes observable.

\begin{table}[ht]
\centering
\begin{tabular}{lll}
\toprule
Quantity & Symbol & Value \\
\midrule
Effective carrier density & $n_{\mathrm{eff}}$ & $1.30\times10^{24}\,\mathrm{m^{-3}}$ \\
Confinement crossover thickness & $L_c$ & $16.9\,\mathrm{nm}$ \\
Bulk mean free path & $\ell_{\rm bulk}$ & $5.0\times10^3\,\mathrm{nm}$ \\
Surface specularity parameter & $p$ & $\simeq 0$ \\
Disorder/grain-boundary parameter & $\gamma$ & $1.21$ \\
Coefficient of determination & $R^2$ & $0.935$ \\
Root-mean-square error & RMSE & $9.13\,\mathrm{nm}$ \\
\bottomrule
\end{tabular}
\caption{
Fit parameters obtained from the comparison with the Al thin-film penetration-depth data. The effective carrier density is taken from the Eliashberg analysis of Ummarino and Zaccone for Al thin films~\cite{Ummarino2025}, which gives a confinement crossover thickness $L_c\simeq16.9\,\mathrm{nm}$.
}
\label{tab:fit_parameters}
\end{table}

The resulting fit gives $\ell_{\rm bulk}\simeq5\,\mu\mathrm{m}$, $p\simeq0$, and $\gamma\simeq1.21$. The nearly vanishing value of $p$ indicates predominantly diffuse surface scattering, as expected for thin evaporated Al films with rough interfaces and native oxide. The value $\gamma$ of order unity indicates that disorder and grain-boundary scattering are comparable in magnitude to the Fuchs--Sondheimer surface-scattering contribution. More generally, the fitted value $\ell_{\rm bulk}\sim5\,\mu\mathrm{m}$ should be interpreted as an effective low-temperature transport length rather than a purely microscopic bulk mean free path. Although large compared with typical room-temperature thin-film values, it remains physically plausible for high-purity Al at cryogenic temperatures where phonon scattering is strongly suppressed. 

The transport parameters entering the fit, namely the bulk mean free
path $\ell_{\rm bulk}$, the surface specularity parameter $p$, and the
phenomenological disorder parameter $\gamma$, primarily affect the
effective mean free path and therefore modify the quantitative magnitude
of the penetration-depth enhancement. These parameters exhibit partial
correlations because different combinations may produce similar values of
$\ell_{\rm eff}(L)$.\\
By contrast, the confinement-induced renormalization of the intrinsic
superconducting coherence length,
\begin{equation}
\xi(L)=\frac{\xi_{\rm bulk}}{\sqrt{C(L)}},
\end{equation}
depends exclusively on the confinement factor $C(L)$ and is therefore
independent of the fitted transport parameters. Consequently, although
the detailed comparison with experiment depends quantitatively on the
chosen transport parameters, the central theoretical prediction of the
present work—the intrinsic confinement-induced renormalization of the
superconducting coherence length—remains robust under reasonable
variations of the fitting parameters.

The decomposition shown in Fig.~\ref{fig:fit} clarifies the physical content of the fit. The direct Ginzburg--Landau confinement contribution modifies the intrinsic coherence length through $\xi(L)=\xi_{\mathrm{bulk}}/\sqrt{C(L)}$, and becomes visible when the reduced thin-film carrier density is used. However, the dominant quantitative increase of the penetration depth is still produced by the reduction of $\ell_{\mathrm{eff}}(L)$ through diffuse surface scattering and disorder. Thus, the data support a physical picture in which confinement renormalizes the intrinsic superconducting length scale, while transport scattering controls the magnitude of the observed penetration-depth enhancement.

Most importantly, however, the present framework goes substantially beyond a conventional Fuchs--Sondheimer description. In the standard FS approach, thickness dependence enters exclusively through the electronic mean free path, while the intrinsic superconducting length scales remain unaffected. In contrast, the theory developed here predicts that quantum confinement directly renormalizes the superconducting coherence length through the confinement-induced modification of the density of states and of the Fermi energy.

In particular, the relation
\begin{equation}
\xi_{\mathrm{film}}(T)=\frac{\xi_{\mathrm{bulk}}(T)}{\sqrt{C(L)}}
\end{equation}
demonstrates that confinement modifies the intrinsic superconducting coherence scale even in the absence of disorder or surface scattering. Furthermore, once disorder effects are included, the effective coherence length becomes
\begin{equation}
\frac{1}{\xi_{\mathrm{film,eff}}(T)}
=
\frac{\sqrt{C(L)}}{\xi_{\mathrm{bulk}}(T)}
+
\frac{1}{C(L)\ell_{\mathrm{bulk}(T)}},
\end{equation}
which explicitly couples quantum confinement and transport scattering into a single superconducting length scale. This confinement-induced renormalization of the coherence length constitutes the central conceptual novelty of the present work.

The comparison with experiment therefore supports the following physical picture. In the experimentally accessible thickness range for Al films, namely tens of nanometers, the dominant contribution to the rapid increase of the penetration depth originates from the suppression of the effective mean free path due to surface and disorder scattering. However, the superconducting response cannot be described purely within a normal-state transport framework because confinement simultaneously modifies the intrinsic superconducting coherence scale through the confinement-dependent density of states. The superconducting electrodynamics of confined films therefore emerges from the interplay between transport scattering and confinement-induced renormalization of the superconducting condensate.

The analysis also demonstrates that a pure Fuchs--Sondheimer description is not quantitatively sufficient to reproduce the experimental upturn of the penetration depth at small thickness. An additional disorder-related scattering channel is required, consistent with the broader physical picture developed recently for confined transport in thin films~\cite{ZacconePRM2025}.

The present comparison should therefore not be interpreted as a direct
experimental proof of the confinement mechanism alone. Rather, it shows
that the confinement-induced renormalization is fully compatible with the
available experimental data and naturally complements the conventional
transport description. In the experimentally accessible thickness range
for Al, the confinement factor remains close to unity, so that transport
scattering dominates the observed enhancement of the penetration depth.
Nevertheless, the theory predicts an intrinsic confinement correction to
the superconducting coherence length that is absent from conventional
surface-scattering models. Future experiments on thinner films or on
materials with larger confinement crossover thicknesses should allow this
intrinsic confinement contribution to be isolated more directly.

\begin{figure}[ht]
\centering
\includegraphics[width=0.75\textwidth]{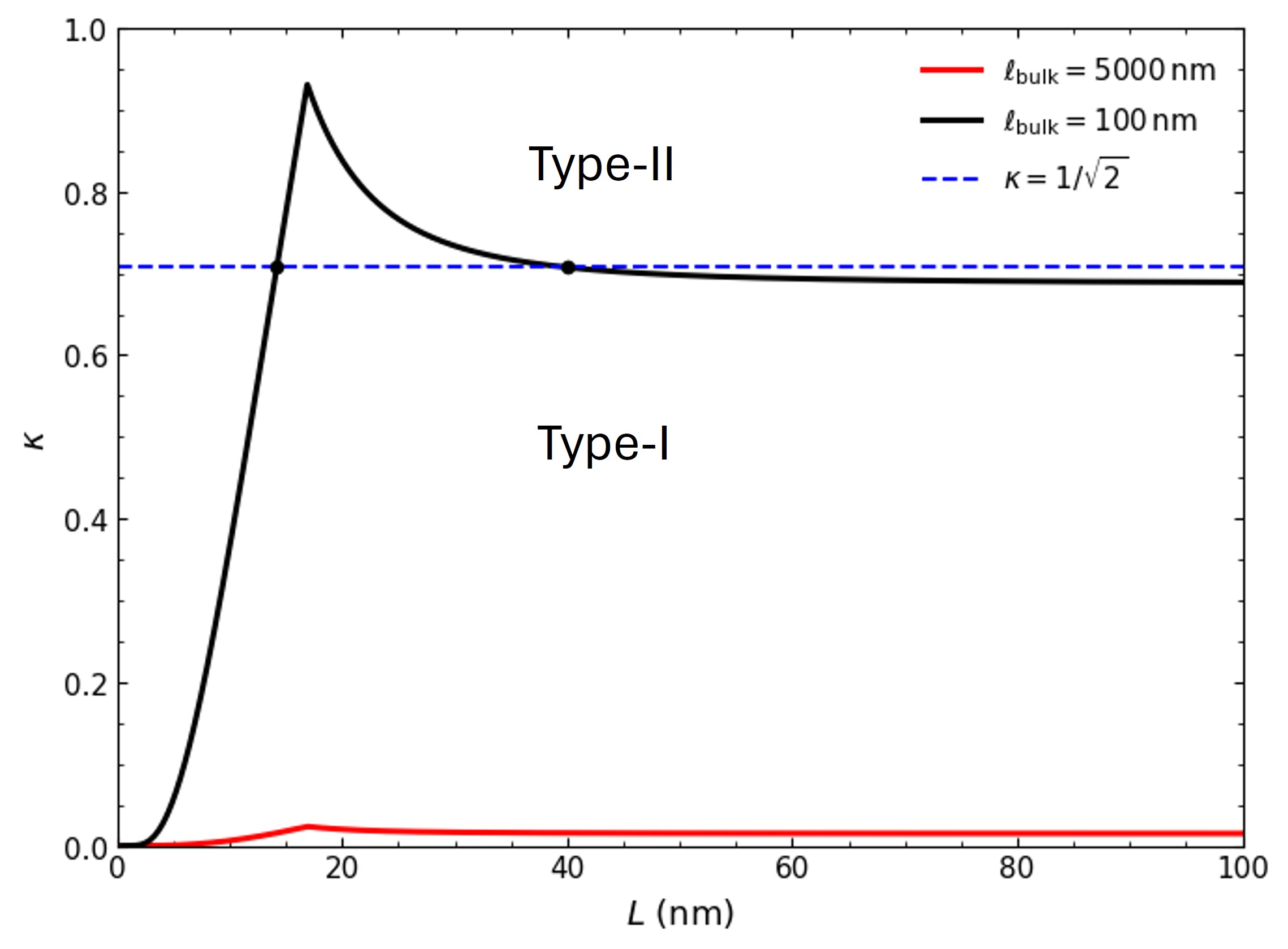}
\caption{
Ginzburg--Landau parameter $\kappa(L)$ calculated from Eq.~(52) as a
function of film thickness. The red curve is obtained using the
parameters extracted from the penetration-depth fit, including
$\ell_{\rm bulk}=5000\,\mathrm{nm}$. The black curve is obtained by
keeping all the remaining parameters fixed while reducing the bulk mean
free path to $\ell_{\rm bulk}=100\,\mathrm{nm}$. The horizontal dashed
line denotes the type-I/type-II boundary,
$\kappa=1/\sqrt{2}$. For the fitted Al parameters, $\kappa(L)$ remains
below the critical value over the whole thickness range considered.
For the shorter mean free path, Eq.~(52) predicts a reentrant
type-I$\rightarrow$type-II$\rightarrow$type-I sequence as the film
thickness is reduced.
}
\label{fig:fit}
\end{figure}

Finally, we investigate the conditions under which a type-I/type-II
crossover may occur in aluminum films. Figure~2 shows the
Ginzburg--Landau parameter $\kappa(L)$ calculated from Eq.~(52) as a
function of film thickness.

The red curve is obtained using the same parameters employed in the
penetration-depth fit, namely
$UN_{\rm bulk}(0)=0.20$,
$\xi_{\rm bulk}=1600\,\mathrm{nm}$,
$\lambda_{L,\rm bulk}=15.7\,\mathrm{nm}$,
$n_{\rm eff}=1.30\times10^{24}\,\mathrm{m}^{-3}$,
and
$\ell_{\rm bulk}=5000\,\mathrm{nm}$.
For these parameters, $\kappa(L)$ increases as the film thickness is
reduced, reflecting the confinement-induced enhancement of type-II
behavior, but it remains below the critical value
$\kappa=1/\sqrt{2}$ throughout the investigated thickness range.
Therefore, no type-I/type-II crossover is predicted for the Al samples
used in the penetration-depth measurements.

To investigate the sensitivity of the crossover to disorder, the black
curve is obtained by keeping all the remaining parameters fixed while
reducing the bulk mean free path to
$\ell_{\rm bulk}=100\,\mathrm{nm}$.
In this case, the stronger dirty-limit correction increases
$\kappa(L)$ sufficiently for the system to cross the
type-I/type-II boundary at approximately
$L\simeq40\,\mathrm{nm}$.
As the thickness is reduced further and the strong-confinement regime is
approached, however, Eq.~(52) predicts a second crossing near
$L\simeq14\,\mathrm{nm}$, where the system returns to the type-I
regime.

This reentrant behavior originates from the competition between the
dirty-limit enhancement of the penetration depth and the suppression of
the superconducting critical temperature in the strong-confinement
regime. At intermediate thicknesses, the dirty-limit contribution
dominates, increasing $\kappa$ and favoring type-II behavior. At very
small thicknesses, however, the rapid confinement-induced reduction of
$T_c$ decreases the London penetration depth sufficiently that
$\kappa$ falls below the critical value again.
The analysis therefore demonstrates that the occurrence and extent of
the type-II regime are strongly sample dependent and controlled by the
combined effects of quantum confinement and disorder through the
electronic mean free path. Within the present phenomenological
framework, strong disorder can induce a finite thickness interval where
the film exhibits type-II superconductivity, bounded on both sides by
type-I behavior.

\section{Conclusions}

We have developed a Ginzburg--Landau framework for confined superconducting thin films based on the recently developed quantum confinement theory of metallic thin films. Explicit analytical expressions were derived for the superconducting coherence length, penetration depth, mean free path, and Ginzburg--Landau parameter in terms of the confinement-induced renormalization of the density of states and Fermi energy.

The central conceptual result of the present work is the prediction that quantum confinement directly renormalizes the intrinsic superconducting coherence length through modifications of the electronic density of states and of the Fermi energy. This effect is fundamentally absent from conventional thin-film transport theories based solely on surface scattering and dirty-limit corrections. The theory therefore predicts that confinement modifies the intrinsic superconducting length scales themselves, even in the absence of disorder or surface scattering.

A particularly important consequence of this mechanism is that confinement simultaneously suppresses the superconducting coherence length while enhancing the penetration depth. As a result, the Ginzburg--Landau parameter increases systematically with decreasing film thickness, implying that quantum confinement intrinsically drives superconductors toward stronger type-II behavior. The interplay between confinement and disorder can even generate a finite thickness interval of type-II superconductivity bounded by type-I phases, illustrating the rich crossover behavior predicted by the present phenomenological framework. This prediction represents a fundamental new consequence of confinement physics in superconductors and provides a direct connection between Fermi-surface reconstruction and macroscopic superconducting electrodynamics.

The theory further predicts a crossover regime in which confinement-induced renormalization of superconducting length scales and transport scattering become strongly intertwined. In this regime, the experimentally observed enhancement of the penetration depth cannot be understood purely within a normal-state transport framework because confinement simultaneously modifies the intrinsic superconducting condensate through the confinement dependence of the coherence length.

Comparison with recent experimental penetration-depth data for Al thin
films shows that the proposed framework consistently describes the
observed evolution of the penetration depth and provides a natural
extension of conventional transport models by incorporating the intrinsic
confinement-induced renormalization of the superconducting coherence
length.

The present framework therefore establishes a unified phenomenological description connecting quantum confinement, Fermi-surface topology, transport scattering, and superconducting electrodynamics in confined metallic films. More broadly, it extends recent confinement theories directly into the Ginzburg–Landau regime and predicts intrinsic confinement-induced modifications of superconducting electrodynamics that should become increasingly important in nanometric superconductors, superconducting quantum devices, and ultrathin superconducting heterostructures.

Future developments should include nonlocal electrodynamics, anisotropic pairing, strong-coupling Eliashberg corrections, explicit grain-size evolution, and fully microscopic treatments of disorder and interface scattering in ultrathin superconductors.
Finally, this methodology could also be applied to studying the effect that the presence of surfaces—and, more generally, boundaries—has on superconductivity \cite{ref1,ref2,ref3,ref4,ref5}.
\\

\section*{Data availability statement}

No new experimental or computational data were generated in this work. The analysis is based exclusively on previously published data cited in the manuscript.\\

\bibliographystyle{iopart-num}
\bibliography{refs}

\end{document}